\documentclass[aps,onecolumn,showpacs,footinbib]{revtex4}%
\usepackage[latin9]{inputenc}
\usepackage{amssymb}
\usepackage{amsmath}
\usepackage{amscd}
\usepackage{graphicx}
\usepackage{subfigure}
\usepackage{float}
\begin{document}
\draft
\title{Dissipation-induced geometric phase for an atom in cavity QED}
\author{Shi-Biao Zheng\thanks{%
E-mail: sbzheng@pub5.fz.fj.cn}}
\address{Department of Physics\\
Fuzhou University\\
Fuzhou 350002, P. R. China}
\date{\today }

\begin{abstract}
We present a feasible scheme to investigate the geometric phase for an atom
trapped in an optical cavity induced by the effective decay process due to
cavity photon loss. The cavity mode, together with the external driving
fields, acts as the engineered environment of the atom. When the parameters
of the reservoir is adiabatically and cyclically changed, the system
initially in the nontrivial dark state of the effective Lindblad operator
undergoes a cyclic evolution and acquires a geometric phase. The geometric
phase can be observed with the atomic Ramsey interference oscillation in the
decoherence-free subspace.
\end{abstract}


\vskip 0.5cm \maketitle

\narrowtext

\section{INTRODUCTION}

In 1984 Berry discovered that a quantum system in a pure state picks up an
extra phase when the parameters of the Hamiltonian make an adiabatic and
cyclic evolution [1,2]. Such a phase, which depends only on the geometry of
the cyclic path traced out by the system and has no time and energy
dependence as compared with the dynamical phase, has been generalized to the
nonadiabatic [3], noncyclic [4], and mixed state [5,6] cases. The Berry
phase produced by unitary evolution has been experimentally tested in
various two-state systems [7-12] and in a resonator [13].

In recent years, the geometric phase has received renewed interest due to
its use in implementation of quantum computation [14-21]. Compared with
schemes relying on dynamical effects, geometric quantum gates have potential
fault tolerance since the geometric phase is robust against any fluctuation
in the control parameters that preserves the area enclosed by the path [22].
For the use in quantum information processing, it is important to understand
the effect of decoherence on the geometric phase due to unavoidable
interaction with the environment. Several papers have investigated the
behavior of geometric phases of a spin-1/2 system [22-26] and a quantum
harmonic oscillator [27,28] in the presence of decoherence. Remarkably,
Carollo et al. showed that the geometric phase of a quantum system can be
generated through slow variation of the parameters of the engineered
reservoir [29]. There exists a decoherence-free subspace (DFS) that is not
affected by interaction with the engineered environment. A state lying in
the DFS is decoupled from states outside the DFS and does not evolve in time
if the reservoir is time independent. However, when the environment depends
upon some parameter that is varied in time, the DFS becomes time dependent.
Under the adiabatic condition, the system initially lying in the DFS
coherently evolves, adiabatically following the DFS. When the DFS undergoes
a cyclic evolution, the system returns to the initial state but acquires a
geometric phase. This is an example of generating parallel transport by
means of an irreversible quantum evolution. To generate this geometric
effect a nontrivial DFS is required. Carollo et al. suggested a way to
observe this effect using a multilevel atom interacting with a broadband
squeezed vacuum with adiabatically changing squeezing parameter [30].
However, how to implement the scheme in a realistic physical system is
experimentally challenging.

In this paper we investigate geometric phases for an atom trapped in an
optical cavity through engineering decay in the framework of cavity QED.
When the cavity decay rate is much larger than the Raman coupling strengths,
the cavity mode can be adiabatically eliminated and the dynamics of the
atomic degree of freedom is modeled by a master equation in Lindblad form.
The effective Lindblad operator, which depends upon the parameters of the
external classical fields, has a nontrivial dark state. When the parameters
of the Lindblad operator are slowly and cyclically varied, this dark state
undergoes a cyclic evolution and acquires a geometric phase. This phase can
be manifested in the interference between this state and the other dark
state that is not affected by the dissipative quantum dynamical process.

\section{THEORETICAL MODEL}

We consider a four-level atoms, having three ground states $\left|
e\right\rangle $, $\left| g\right\rangle $, and $\left| f\right\rangle $ and
an excited state $\left| r\right\rangle $, trapped in a single-mode optical
cavity. The transition $\left| e\right\rangle \rightarrow \left|
r\right\rangle $ ($\left| f\right\rangle \rightarrow \left| r\right\rangle $%
) is driven by two classical laser fields with the same Rabi frequency $%
\Omega _1$ ($\Omega _2$) and phase $\varphi _1$ ($\varphi _2$), but with
opposite detunings $\Delta $ and $-\Delta $, as shown in Fig. 1. The
transition $\left| g\right\rangle \rightarrow \left| r\right\rangle $ is
coupled to the cavity mode with the coupling constant $g$ and detuning $%
\Delta $. In the interaction picture, the Hamiltonian is
\begin{eqnarray}
H &=&gae^{i\Delta t}\left| r\right\rangle \left\langle g\right| +\Omega
_1[e^{i(\Delta t+\varphi _1)}+e^{i(-\Delta t+\varphi _1)}]\left|
r\right\rangle \left\langle e\right|  \nonumber  \label{3} \\
&&\ +\Omega _2[e^{i(\Delta t+\varphi _2)}+e^{i(-\Delta t+\varphi _2)}]\left|
r\right\rangle \left\langle f\right| +H.c.,
\end{eqnarray}
where $a$ is the annihilation operator of the cavity mode. Under the
condition $\Delta \gg \Omega _j,g$ the upper level $\left| r\right\rangle $
can be adiabatically eliminated. The Raman couplings and Stark shifts
induced by the two pairs of classical fields cancel out due to the opposite
detunings. On the other hand, the cavity mode, together with the classical
field coupled to $\left| e\right\rangle \rightarrow \left| r\right\rangle $ (%
$\left| f\right\rangle \rightarrow \left| r\right\rangle $) with the
detuning $\Delta $, induces the Raman transition between the ground states $%
\left| g\right\rangle $ and $\left| e\right\rangle $ ($\left| f\right\rangle
$). Then the effective Hamiltonian is given by
\begin{equation}
H_e=\frac{g^2}\Delta a^{+}a\left| g\right\rangle \left\langle g\right|
+[\lambda _1e^{-i\varphi _1}a\left| e\right\rangle \left\langle g\right|
+\lambda _2e^{-i\varphi _2}a\left| f\right\rangle \left\langle g\right|
)+H.c.],
\end{equation}
where $\lambda _j=\frac{\Omega _jg}\Delta $.

In the limit $\kappa \gg $ $\frac{g^2}\Delta ,\lambda _j$, where $\kappa $
is the cavity decay rate, the cavity mode is only virtually populated and
can be adiabatically eliminated. Then the dynamics of the atom is described
by the master equation
\begin{equation}
\stackrel{.}{\rho }=2L\rho L^{\dagger }-\rho L^{\dagger }L-L^{\dagger }L\rho
,
\end{equation}
where $L=\sqrt{\Gamma }(\sin \frac \theta 2\left| g\right\rangle
\left\langle e\right| +e^{-i\varphi }\cos \frac \theta 2\left|
g\right\rangle \left\langle f\right| )$, $\Gamma =\lambda ^2/\kappa $, $%
\lambda =[\lambda _1^2+\lambda _2]^{1/2}$, $\sin \frac \theta 2=\lambda
_1/\lambda $, and $\varphi =\varphi _1-\varphi _2$. It can be easily shown
that the state $\left| \psi _d\right\rangle =\cos \frac \theta 2\left|
e\right\rangle -e^{i\varphi }\sin \frac \theta 2\left| f\right\rangle $ is
the dark state of the engineered Lindblad operator $L$, i.e., $L\left| \psi
_d\right\rangle =0$. In terms of the basis \{$\left| e\right\rangle ,\left|
f\right\rangle ,\left| g\right\rangle $\} $\left| \psi _d\right\rangle $ can
be expressed as $\left| \psi _d\right\rangle =(\cos \frac \theta 2%
,-e^{i\varphi }\sin \frac \theta 2,0)^T$. This state and the trivial ground
state $\left| g\right\rangle $ form the DFS. States lying in such a subspace
remain unaffected by the dissipative quantum dynamical process induced by
cavity loss. The bright state of the engineered Lindblad term $L^{\dagger }L$
is $\left| \psi _b\right\rangle =(e^{-i\varphi }\sin \frac \theta 2,\cos
\frac \theta 2,0)^T$. Such a state would decay to the ground state $\left|
g\right\rangle $ due to the engineered dissipative process. This model is
distinguished from that described in Ref. [30] in which there is only one
decoherence-free state and all of the other states coupled to the squeezed
vacuum reservoir would decay to it.

\section{DISSIPATION-INDUCED GEOMETRIC PHASE}

Since the state $\left| \psi _d\right\rangle $ depends upon the parameters
of the Lindblad operator $L$, one can adiabatically follow the dark state in
the DFS by slowly changing the parameters $\theta $ and $\varphi $ of $L$.
On one hand, the adiabatic manipulation of the decay parameters leads to the
parallel transport of the dark state $\left| \psi _d\right\rangle $. On the
other hand, the change of the dark state in time causes the transition $%
\left| \psi _d\right\rangle \rightarrow $ $\left| \psi _b\right\rangle $
followed by a rapid decay to the ground state $\left| g\right\rangle $. The
transition to the state $\left| g\right\rangle $ is allowed, but the
transition away from it is forbidden since it is not pumped. Therefore, $%
\left| g\right\rangle $ is the steady state, whose population is increased
during the variation of the decay parameters. However, when the change rate
is small enough the leakage probability from $\left| \psi _d\right\rangle $
to $\left| g\right\rangle $ through $\left| \psi _b\right\rangle $ is
negligible and the pure effect is the adiabatic coherent evolution of the
nontrivial dark state in the DFS. It should be noted that the states $\left|
e\right\rangle $ and $\left| f\right\rangle $ have a nonvanishing
probability to be excited by the pump fields during the adiabatic evolution.
In experimental realizations, one must consider the influence of this
excitation as a source of systematic errors. After a cyclic evolution of the
parameters, the dark state would traverse a circuit in the projected Hilbert
space and gain a geometric phase. Unlike the usual Berry's adiabatic phase,
this phase is associated with the incoherent process, other than the unitary
dynamics. The parallel transport is generated via decay engineering. To show
how the adiabatic manipulation of resevior can take place, consider the
unitary transformation
\begin{equation}
U=\left(
\begin{array}{ccc}
\cos \frac \theta 2e^{i\varphi /2} & -\sin \frac \theta 2e^{-i\varphi /2} & 0
\\
\sin \frac \theta 2e^{i\varphi /2} & \cos \frac \theta 2e^{-i\varphi /2} & 0
\\
0 & 0 & 1
\end{array}
\right) .
\end{equation}
Under this change the states $\left| \psi _d\right\rangle $ and $\left| \psi
_b\right\rangle $ are transformed to $(e^{i\varphi /2},0,0)^T$ and $%
(0,e^{i\varphi /2},0)^T$, respectively. On the other hand, $\left|
g\right\rangle $ remains unchanged, i.e., $U\left| g\right\rangle =\left|
g^{^{\prime }}\right\rangle =(0,0,1)^T$. This transformation corresponds to
a reference frame whose basis states $\left| e^{^{\prime }}\right\rangle
=(1,0,0)^T$ and $\left| f^{^{\prime }}\right\rangle =(0,1,0)^T$ respectively
coincide with the time-dependent states $e^{-i\varphi /2}\left| \psi
_d\right\rangle $ and $e^{-i\varphi /2}\left| \psi _b\right\rangle $ in the
original frame, and the geometric effect arising from the cyclic evolution
of the engineered reservoir parameters is manifested in the phase change of
the coherence between the two states $\left| e^{^{\prime }}\right\rangle $
and $\left| g^{^{\prime }}\right\rangle $. After the transformation the
master equation becomes
\begin{equation}
\stackrel{.}{\rho }^{^{\prime }}=2L^{^{\prime }}\rho ^{^{\prime
}}L^{^{\prime }\dagger }-\rho ^{^{\prime }}L^{^{\prime }\dagger }L^{^{\prime
}}-L^{^{\prime }\dagger }L^{^{\prime }}\rho ^{^{\prime }}+\stackrel{.}{U}%
U^{\dagger }\rho ^{^{\prime }}+\rho ^{^{\prime }}U\stackrel{.}{U}^{\dagger },
\end{equation}
where $\rho ^{^{\prime }}=U\rho U^{\dagger }$ and $L^{^{\prime
}}=ULU^{\dagger }$. In this frame the Lindblad term $L^{^{\prime }\dagger
}L^{^{\prime }}$ has the diagonal form:

\begin{equation}
L^{^{\prime }\dagger }L^{^{\prime }}=\Gamma \left| f^{^{\prime
}}\right\rangle \left\langle f^{^{\prime }}\right| .
\end{equation}
The last two terms of the master equation arise from the time-dependent
unitary transformation which is determined by the decay parameters. In the
rotating frame the three eigenstates of the Lindblad operator $L^{^{\prime
}\dagger }L^{^{\prime }}$ corresponds to the three basis states so that it
is convenient to calculate the evolution of the coherence between these
states.

Suppose that the parameter $\theta $ is kept constant and $\phi $ is slowly
changed from $0$ to $2\pi $. In terms of the basis \{$\left| e^{^{\prime
}}\right\rangle ,$ $\left| f^{^{\prime }}\right\rangle ,\left| g^{^{\prime
}}\right\rangle $\} in the rotating frame the evolution of the system is
described by the following coupled differential equations for the density
matrix elements:

\begin{eqnarray}
\stackrel{.}{\rho }_{e,e}^{^{\prime }} &=&\frac i2\sin \theta \stackrel{.}{%
\varphi }(\rho _{f,e}^{^{\prime }}-\rho _{e,f}^{^{\prime }}),  \nonumber \\
\stackrel{.}{\rho }_{g,g}^{^{\prime }} &=&2\Gamma \rho _{f,f}^{^{\prime }},
\nonumber \\
\stackrel{.}{\rho }_{f,f}^{^{\prime }} &=&-2\Gamma \rho _{f,f}^{^{\prime }}+%
\frac i2\sin \theta \stackrel{.}{\varphi }(\rho _{e,f}^{^{\prime }}-\rho
_{f,e}^{^{\prime }}),  \nonumber \\
\stackrel{.}{\rho }_{e,g}^{^{\prime }} &=&\frac i2\stackrel{.}{\varphi }%
(\cos \theta \rho _{e,g}^{^{\prime }}+\sin \theta \rho _{f,g}^{^{\prime }}),
\nonumber \\
\stackrel{.}{\rho }_{f,g}^{^{\prime }} &=&-\Gamma \rho _{f,g}^{^{\prime }}+%
\frac i2\stackrel{.}{\varphi }(\sin \theta \rho _{e,g}^{^{\prime }}-\cos
\theta \rho _{f,g}^{^{\prime }}),  \nonumber \\
\stackrel{.}{\rho }_{e,f}^{^{\prime }} &=&-\Gamma \rho _{e,f}^{^{\prime }}+%
\frac i2\sin \theta \stackrel{.}{\varphi }(-\rho _{e,e}^{^{\prime }}+\rho
_{f,f}^{^{\prime }})+i\cos \theta \stackrel{.}{\varphi }\rho
_{e,f}^{^{\prime }},
\end{eqnarray}
where $\rho _{j,k}^{^{\prime }}=\left\langle j^{^{\prime }}\right| \rho
^{^{\prime }}\left| k^{^{\prime }}\right\rangle $ ($j,k=e,f,g$). The motion
of the matrix elements $\rho _{e,g}^{^{\prime }}(t)$ and $\rho
_{f,g}^{^{\prime }}(t)$ are decoupled from other elements. Suppose that $%
\rho _{f,g}^{^{\prime }}(0)=0$. Then we have
\begin{eqnarray}
\rho _{e,g}^{^{\prime }}(t) &=&\frac{\rho _{e,g}^{^{\prime }}(0)}{\lambda
_{-}-\lambda _{+}}[(\lambda _{-}-\frac i2\cos \theta \stackrel{.}{\varphi }%
)e^{\lambda _{+}t}-(\lambda _{+}-\frac i2\cos \theta \stackrel{.}{\varphi }%
)e^{\lambda _{-}t}],  \nonumber \\
\rho _{f,g}^{^{\prime }}(t) &=&\frac{2\rho _{e,g}^{^{\prime }}(0)}{i(\lambda
_{-}-\lambda _{+})\sin \theta \stackrel{.}{\varphi }}(\lambda _{-}-\frac i2%
\cos \theta \stackrel{.}{\varphi })(\lambda _{+}-\frac i2\cos \theta
\stackrel{.}{\varphi })(e^{\lambda _{+}t}-e^{\lambda _{-}t}),
\end{eqnarray}
where $\lambda _{\pm }=\frac 12(-\Gamma \pm \sqrt{\Gamma ^2+2i\Gamma \cos
\theta \stackrel{.}{\varphi }-\stackrel{.}{\varphi }^2})$. We here have
assumed that $\stackrel{.}{\varphi }$ is a constant. Under the condition $%
\stackrel{.}{\varphi }\ll \Gamma $ we can retain these matrix elements to
the first order in $\stackrel{.}{\varphi }/\Gamma $. At the time $T=2\pi /%
\stackrel{.}{\varphi }$ when the parameter $\varphi $ makes a cyclic
evolution, these two matrix elements are approximately given by
\begin{eqnarray}
\rho _{e,g}^{^{\prime }}(T) &\simeq &\rho _{e,g}^{^{\prime }}(0)e^{i\pi \cos
\theta -\pi \sin ^2\theta \stackrel{.}{\varphi }/2\Gamma },  \nonumber \\
\rho _{f,g}^{^{\prime }}(T) &\simeq &\rho _{e,g}^{^{\prime }}(0)\frac{i\sin
\theta \stackrel{.}{\varphi }}{2\Gamma }e^{i\pi \cos \theta -\pi \sin
^2\theta \stackrel{.}{\varphi }/2\Gamma }.
\end{eqnarray}
We here have discarded the terms decaying at the rate $\Gamma $. This result
shows that the coherence $\rho _{e,g}^{^{\prime }}(t)$ is well preserved
during the engineered dissipative process. To the first order in $\stackrel{.%
}{\varphi }/\Gamma $ the other matrix elements are
\begin{eqnarray}
\rho _{e,e}^{^{\prime }}(T) &\simeq &(1-\frac{\pi \sin ^2\theta \stackrel{.}{%
\varphi }}\Gamma )\rho _{e,e}^{^{\prime }}(0),  \nonumber \\
\rho _{g,g}^{^{\prime }}(T) &\simeq &\rho _{g,g}^{^{\prime }}(0)+\frac{\pi
\sin ^2\theta \stackrel{.}{\varphi }}\Gamma \rho _{e,e}^{^{\prime }}(0)
\nonumber \\
\rho _{f,f}^{^{\prime }}(T) &\simeq &0,  \nonumber \\
\rho _{e,f}^{^{\prime }}(T) &\simeq &\frac{-i\sin \theta \stackrel{.}{%
\varphi }}{2\Gamma }\rho _{e,e}^{^{\prime }}(0).
\end{eqnarray}

Reversing the unitary transformation $U(T)$ we obtain the density matrix
elements in the original frame
\begin{eqnarray}
\rho _{d,d}(T) &\simeq &(1-\frac{\pi \sin ^2\theta \stackrel{.}{\varphi }}%
\Gamma )\rho _{d,d}(0),\text{ }  \nonumber \\
\rho _{d,b}(T) &\simeq &\frac{-i\sin \theta \stackrel{.}{\varphi }}{2\Gamma }%
\rho _{d,d}(0),  \nonumber \\
\rho _{b,b}(T) &\simeq &0,\text{ }  \nonumber \\
\rho _{g,g}(T) &=&\rho _{g,g}(0)+\frac{\pi \sin ^2\theta \stackrel{.}{%
\varphi }}\Gamma \rho _{d,d}(0),  \nonumber \\
\rho _{d,g}(T) &\simeq &\rho _{d,g}(0)e^{i\beta -\pi \sin ^2\theta \stackrel{%
.}{\varphi }/2\Gamma },  \nonumber \\
\rho _{b,g}(T) &\simeq &\rho _{d,g}(0)\frac{i\sin \theta \stackrel{.}{%
\varphi }}{2\Gamma }e^{i\beta -\pi \sin ^2\theta \stackrel{.}{\varphi }%
/2\Gamma },
\end{eqnarray}
where $\beta =(\cos \theta -1)\pi $, $\rho _{g,g}(T)=\left\langle g\right|
\rho (T)\left| g\right\rangle $, $\rho _{j,g}(T)=\left\langle \psi _j\right|
\rho (T)\left| g\right\rangle $, and $\rho _{j,k}(T)=\left\langle \psi
_j\right| \rho (T)\left| \psi _k\right\rangle $ ($j,k=d,b$). The geometric
phase induced by the steering process is encoded in the phase of the
coherence $\rho _{d,g}(T)$ in the DFS. The adiabatic manipulation of the
decay dynamics drives the atom to undergo a parallel transport if it is
initially in the dark state $\left| \psi _d\right\rangle $. Due to the time
evolution of the DFS, the atom has a probability of leaking out of this
subspace to the bright state $\left| \psi _b\right\rangle $, which rapidly
decays to $\left| g\right\rangle $. This causes the damp of the coherence in
the DFS. In the limit $\stackrel{.}{\varphi }\ll \Gamma $ the main effect of
the incoherent process is just a phase factor $e^{i\beta }$. Since the state
$\left| g\right\rangle $ is not affected by the steering process, the phase $%
\beta $ is associated with the cyclic evolution of the dark state $\left|
\psi _d\right\rangle $ due to adiabatic variation of the parameters of the
Lindblad operator. The acquired phase also has a geometric interpretation in
the parameter space of the Lindblad operator $L$: $\beta =-\Theta /2$, where
$\Theta $ is the solid swept by the vector always pointing to $(\theta
,\varphi )$ on the Poincar\'e sphere. It is worth noting that this phase
coincides with the usual Berry geometric phase acquired by the dark state $%
\left| \psi _d\right\rangle \left| 0\right\rangle $ of the Hamiltonian (2)
when the Hamiltonian parameters are dragged along a closed loop in the
absence of decoherence, where $\left| 0\right\rangle $ is the vacuum state
of the cavity mode. This is because in both processes the dark state
traverses the same closed path in the projected Hilbert space as long as the
change rate of the parameters is sufficiently slow.

\section{MEASUREMENT OF THE GEOMETRIC PHASE}

Now we show that such a phase can be measured by the Ramsey interference.
Suppose that the atom is initially prepared in the superposition state $%
\frac 1{\sqrt{2}}(\left| \psi _d(0)\right\rangle +\left| g\right\rangle )$.
After the cyclic evolution of the Lindblad operator the state of the system
is
\begin{eqnarray}
\rho  &\simeq &(1-\frac{\pi \sin ^2\theta \stackrel{.}{\varphi }}{2\Gamma }%
)\left| \phi \right\rangle \left\langle \phi \right| +\frac{3\pi \sin
^2\theta \stackrel{.}{\varphi }}{4\Gamma }\left| g\right\rangle \left\langle
g\right|   \nonumber \\
&&\ -\frac{\pi \sin ^2\theta \stackrel{.}{\varphi }}{4\Gamma }\left| \psi
_d(0)\right\rangle \left\langle \psi _d(0)\right| ,
\end{eqnarray}
where
\begin{equation}
\left| \phi \right\rangle =\frac 1{\sqrt{2+(\frac{\sin \theta \stackrel{.}{%
\varphi }}{2\Gamma })^2}}[e^{i\beta }\left| \psi _d(0)\right\rangle +\left|
g\right\rangle +\frac{\sin \theta \stackrel{.}{\varphi }}{2\Gamma }%
e^{i\alpha }\left| \psi _b(0)\right\rangle ]
\end{equation}
and $\alpha =(\cos \theta +3/2)\pi $. After the steering process, we apply
two classical pulses to the atom. The first classical pulse produces the
transformation $\left| e\right\rangle \rightarrow \cos \frac \theta 2\left|
e\right\rangle +\sin \frac \theta 2\left| f\right\rangle $ and $\left|
f\right\rangle \rightarrow \cos \frac \theta 2\left| f\right\rangle -\sin
\frac \theta 2\left| e\right\rangle $, while the second leads to $\left|
g\right\rangle \rightarrow \frac 1{\sqrt{2}}(\left| g\right\rangle +\left|
e\right\rangle )$ and $\left| e\right\rangle \rightarrow \frac 1{\sqrt{2}}%
(\left| e\right\rangle -\left| g\right\rangle )$. Then the state of the atom
evolves to
\begin{eqnarray}
\rho ^{^{\prime }} &\simeq &(1-\frac{\pi \sin ^2\theta \stackrel{.}{\varphi }%
}{2\Gamma })\left| \phi ^{^{\prime }}\right\rangle \left\langle \phi
^{^{\prime }}\right| +\frac{3\pi \sin ^2\theta \stackrel{.}{\varphi }}{%
8\Gamma }(\left| g\right\rangle +\left| e\right\rangle )(\left\langle
g\right| +\left\langle e\right| )  \nonumber \\
&&\ -\frac{\pi \sin ^2\theta \stackrel{.}{\varphi }}{8\Gamma }(\left|
e\right\rangle -\left| g\right\rangle )(\left\langle e\right| -\left\langle
g\right| ),
\end{eqnarray}
where
\begin{equation}
\left| \phi ^{^{\prime }}\right\rangle =\frac 1{\sqrt{2+(\frac{\sin \theta
\stackrel{.}{\varphi }}{2\Gamma })^2}}[e^{i\beta }(\left| e\right\rangle
-\left| g\right\rangle )/\sqrt{2}+(\left| g\right\rangle +\left|
e\right\rangle )/\sqrt{2}+\frac{\sin \theta \stackrel{.}{\varphi }}{2\Gamma }%
e^{i\alpha }\left| f\right\rangle ].
\end{equation}
Finally, the probabilities of finding the atom in the states $\left|
g\right\rangle $ and $\left| e\right\rangle $ are given by
\begin{equation}
P_{g,e}\simeq \frac 12[1\mp (1-\frac{\pi \sin ^2\theta \stackrel{.}{\varphi }%
}{2\Gamma })\cos \beta ],
\end{equation}
which is independent of the atom-field interaction time. By varying the
geometric phase $\beta $ the probability of finding the atom in the state $%
\left| g\right\rangle $ or $\left| e\right\rangle $ exhibits Ramsey
interference fringes, with the visibility shrunk by a factor
\begin{equation}
V=1-\frac{\pi \sin ^2\theta \stackrel{.}{\varphi }}{2\Gamma }.
\end{equation}
It should be noted that the reduction of the visibility is unavoidable since
the acquired geometric phase is associated with the incoherent dynamics.

We proceed to consider the effect of atomic spontaneous emission. The
classical fields do not induce the atomic transition from the dark state $%
\left| \psi _d\right\rangle $ to the excited state $\left| r\right\rangle $
due to destructive quantum interference between the transition paths. The
atomic spontaneous emission arises from the coupling of the state $\left|
\psi _b\right\rangle $ to the excited state with the effective decay rate $%
\gamma _e\sim \gamma P_b\Omega ^2/\Delta ^2$, where $P_b=$ $\frac{\sin
^2\theta \stackrel{.}{\varphi }^2}{8\Gamma ^2}$ and $\gamma $ is the atomic
spontaneous emission rate. This further reduces the atomic coherence by a
factor $V^{^{\prime }}\sim e^{-\gamma _eT}$. The required atomic level
configuration can be achieved in Cs. The Zeeman sublevels $\left|
F=4,m=-4\right\rangle $, $\left| F=4,m=-3\right\rangle $, and $\left|
F=3,m=-3\right\rangle $ of $6^2S_{1/2}$ can act as the ground states $\left|
g\right\rangle $, $\left| e\right\rangle $, and $\left| f\right\rangle $,
respectively, while the Zeeman sublevel $\left| F^{^{\prime }}=4,m^{^{\prime
}}=-4\right\rangle $ of $6^2P_{3/2}$ can act as the excited state $\left|
e\right\rangle $. In cavity QED experiments with Cs atoms trapped in an
optical cavity, the corresponding atom-cavity coupling strength is $g=2\pi
\times 34$ MHz [31]. The decay rates for the atomic excited state and the
cavity mode are $\gamma =2\pi \times 2.6$ MHz and $\kappa =2\pi \times 4.1$
MHz, respectively. Setting $\Omega =g$, $\Delta =100g$, $\theta =\pi /4$,
and $\stackrel{.}{\varphi }=10^{-4}g$, we have $\Gamma \simeq $ $2\pi \times
0.0282$ MHz, $V\simeq 0.91$, and $V^{^{\prime }}\sim e^{-4\times 10^{-4}}$,
which means that the the Ramsey oscillation is almost not affected by the
atomic spontaneous emission and the visibility of the interference fringes
is about 91\%.

\section{CONCLUSION}

In conclusion, we have proposed a scheme for generating and measuring
geometric phases based on reservoir engineering in the framework of cavity
QED. The decaying cavity mode and the external driving fields form the
engineered environment for the atom. By adiabatically changing the
environment parameters the atom can be driven to undergo a coherent cyclic
evolution in DFS, picking up a phase that depends upon the geometry of the
path executed in the parameter space of the engineered Lindblad operator.
The acquired phase is purely geometrical since there is no dynamical
evolution during the manipulation of the engineered environment. This phase
can be measured through the Ramsey interference in DFS. Based on cavity QED
techniques presently or soon to be available the proposed scheme might be
realizable.

This work was supported by the Major State Basic Research Development
Program of China under Grant No. 2012CB921601, the National Natural Science
Foundation of China under Grant No. 10974028, the Doctoral Foundation of the
Ministry of Education of China under Grant No. 20093514110009, and the
Natural Science Foundation of Fujian Province under Grant No. 2009J06002.

Fig. 1 (color online). The atomic level configuration and excitation scheme.
The transition $\left| e\right\rangle \rightarrow \left| r\right\rangle $ ($%
\left| f\right\rangle \rightarrow \left| r\right\rangle $) is driven by two
classical laser fields with the same Rabi frequency $\Omega _1$ ($\Omega _2$%
) and phase $\varphi _1$ ($\varphi _2$) but with opposite detunings
$\Delta $ and $-\Delta $. The transition $\left| g\right\rangle
\rightarrow \left| r\right\rangle $ is coupled to the cavity mode
with the coupling constant $g$ and detuning $\Delta
$.
\begin{figure}[C]
\includegraphics[width=0.5\columnwidth]{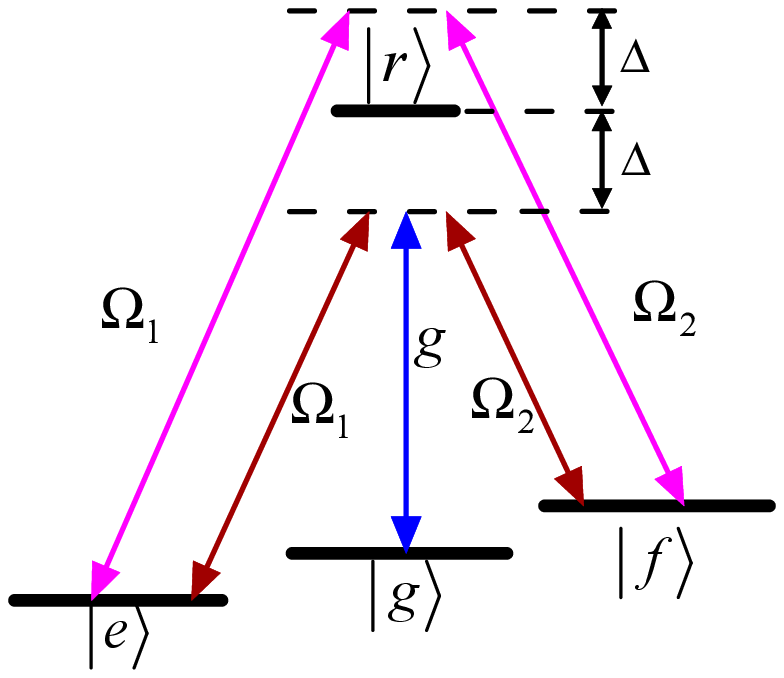}
\caption{}
\end{figure}

\end{document}